# Phonon-polaritons in Zn$_{1-x}$Mg$_x$Te ($x \leq 0.09$): A Raman scattering study

D. Singh,[1,2] A. Elmahjoubi,[1] O. Pagès,[1,a)] V.J.B. Torres,[3] C. Gardiennet,[4] G. Kervern,[4] A. Polian,[5] Y. Le Godec,[5] J.-P. Itié,[6] S. Diliberto,[7] S. Michel,[7] P. Franchetti,[1] and K. Strzałkowski[2]

1. Université de Lorraine, LCP-A2MC, UR 201019679B, F-57000 Metz, France
2. Institute of Physics, Faculty of Physics, Astronomy and Informatics, Nicolaus Copernicus University in Toruń, ul. Grudziądzka 5, 87-100 Toruń, Poland
3. Departamento de Fisica and I3N, Universidade de Aveiro, 3810 – 193 Aveiro, Portugal Institut de
4. Université de Lorraine, Laboratoire de Cristallographie, Résonance Magnétique et Modélisations, UMR 7036, Vandoeuvre-lès-Nancy, F-54506, France
5. Institut de Minéralogie, de Physique des Matériaux et de Cosmochimie, Sorbonne Université — UMR CNRS 7590, F-75005 Paris, France
6. Synchrotron SOLEIL, L'Orme des Merisiers Saint-Aubin, BP 48 F-91192 Gif-sur-Yvette Cedex, France
7. Université de Lorraine, CNRS, IJL, F-57000, Metz, France

a) Author to whom correspondence should be addressed : olivier.pages@univ-lorraine.fr

**Abstract.** Phonon-polaritons (PP) are phonon-photon coupled modes. Using near-forward Raman scattering, the PP of the cubic Zn$_{1-x}$Mg$_x$Te ($x \leq 0.09$) semiconductor alloy could be measured. While the PP-coupling hardly develops in pure ZnTe, minor Mg-alloying suffices to stabilize a long-lifetime PP strongly bound to the lattice, *i.e.*, with a pronounced phonon character, and yet a fast one originating from the highly dispersive photon-like bottleneck of the PP-dispersion. By combining the advantages of a phonon and of a photon, the long-lifetime PP generated by minor Mg-alloying of ZnTe marks an improvement over the PP of pristine ZnTe, that, from the Raman cross section calculation, can only achieve a balanced compromise between the two kinds of advantages, intensity and speed. The discussion of the PP-related lattice dynamics of Zn$_{1-x}$Mg$_x$Te ($x \leq 0.09$) is grounded in a preliminary study of the lattice macro- and microstructure using X-ray diffraction and solid-state nuclear magnetic resonance, respectively, and further relies on *ab initio* calculations of the native phonon modes behind the PP in the Mg-dilute limit of Zn$_{1-x}$Mg$_x$Te ($x \sim 0$), considering various Mg-isotopes.

Due to the ionicity of the cation-anion chemical bonding in cubic (zincblende) semiconductor compounds like ZnTe, the optical lattice vibrations (phonons), involving opposite displacement of the cationic (Zn) and anionic (Te) species, carry an electric field. The transverse optical (TO) mode, referring to atom vibrations perpendicular to the direction of propagation, is interesting for photonics in that its electric field is transverse, *i.e.*, photon-like. The resulting TO excitation with mixed phonon-photon character, the so-called phonon-polariton (PP),[1] offers the possibility to transfer information throughout matter (phonon character) at high velocity (photon character) in the terahertz range via lattice vibrations.

However, a TO mode can propagate as a PP only if its wavevector $\vec{q}$ remains compatible with the existence of a photon, *i.e.*, a pure transverse electric field. The condition is that $q$ falls near the quasi vertical $\omega(q)$ dispersion of a photon, governed by the speed of light in vacuum scaled down by the refractive index of the crystal. Otherwise the TO mode is deprived of electric field and reduces to a purely mechanical TO (PM-TO) phonon. Experimentally, the PP-coupling can be observed by Raman scattering by adopting the forward scattering geometry (schematically operated "in transmission"). This achieves minimal $\vec{q}$-transfer to the crystal, of the order of $10^{-4}$ of the Brillouin zone size.[1-3] In the classical backward scattering geometry, the $\vec{q}$-transfer is maximum, of the order of $10^{-2}$ of the Brillouin zone size. The TO mode is then probed in its asymptotic PM-TO regime as a pure phonon.[1-3]

One major drawback of PP propagating in volume of pristine zincblende crystals is that, generally, they become hardly supported by the crystal as soon as they depart from their native PM-TO phonon and engage the deep photon-like regime, as testified by a rapid collapse of the PP Raman cross section.[1,2] Such collapse conforms to intuition, verified experimentally,[4] that only matter/phonon-like excitations scatter light efficiently, not photon-like ones. One thus faces a serious dilemma with the PP of pristine crystals. Either they are well supported by the lattice (phonon-like) but slow, as observed, *e.g.*, in ZnSe (Ref.[4], Fig. S2 therein) and ZnTe (Ref.[5], Fig. S3 therein), or they are fast (photon-like) but hardly supported by the lattice. The advantages of a phonon and of a photon seem mutually exclusive in pristine crystals. One can only achieve a balanced compromise.

$A_{1-x}B_xC$-alloying may avoid the need to compromise. *A priori*, alloying is detrimental to PP propagation since the A↔B substitution generates mass and force-constant disorders, aggravated by local lattice distortions around the A↔B substitutional sites. Altogether, these represent major obstacles to the lattice dynamics. Probably this is the reason why the alloy-related PP remain unexplored experimentally in the literature – apart in our own recent works. Yet, intense (hence phonon-like) and sharp (reflecting a long-lifetime) Raman signals stemming from the deep photon-like PP regime have lately been observed after moderate-to-large alloying of ZnSe by substituting the light Be ($Zn_{0.47}Be_{0.53}Se$)[6] and Mg ($Zn_{0.74}Mg_{0.26}Se$)[4] elements for Zn and the light S element for Se ($ZnSe_{0.68}S_{0.32}$)[7]. Similar attempts by minor-to-moderate substitution of the heavy Cd element for Zn in either ZnSe ($Cd_{0.075}Zn_{0.925}Se$)[8] or ZnTe ($Cd_{0.09}Zn_{0.91}Te$)[5] fell short of achieving the deep photon-like regime. Only the shallow phonon-like (PM-TO) regime could be probed.

Spurred on by the initial success with the light substituents, we find motivation in this work to explore whether the PP-coupling in volume of ZnTe, so far unsuccessful,[5] can be improved by minor Mg-alloying. This issue is further interesting *per se* since the PP modes of $Zn_{1-x}Mg_xTe$ are unexplored. The native $Zn_{1-x}Mg_xTe$ phonons behind the PP, however, are well-documented.[9-11] Two clearly separated ($\Delta\omega \sim 80$ cm$^{-1}$) Zn-Te and Mg-Te PM-TO modes are identified across the composition domain, as explained by the historical modified-random-element-isodisplacement (MREI) model.[12]

We report below on a near-forward Raman study of the $Zn_{1-x}Mg_xTe$-related PP at minor Mg-alloying ($x \leq 0.09$) using red 632.8 nm and near-infrared 785 nm laser excitations. Due to its large optical band gap at minor Mg content ($E_g \sim 2.30$ eV at 300 K)[13] $Zn_{1-x}Mg_xTe$ is transparent to such radiations, a prerequisite for near-forward (transmission-like) Raman measurements. The near-forward Raman study of the PP is grounded in a preliminary backward Raman study of their native Zn-Te and Mg-Te PM-TO modes. The latter is hardly observed experimentally. Its assignment is supported by *ab initio* calculations of the local vibration modes due a to a minimal Mg impurity motif in ZnTe using the AIMPRO[14,15] code (Ab Initio Modeling PROgram), considering various stable Mg isotopes. The Zn-Te and Mg-Te PM-TO frequencies enter as main input parameters into the $Zn_{1-x}Mg_xTe$ PP Raman cross section[5] providing an overview of the PP Raman signal in its $(q, x)$-dependence, useful to discuss the experimental PP data.

The used $Zn_{1-x}Mg_xTe$ samples are grown by the Bridgman method[16] at small Mg content ($x$=0, 0.023, 0.046, 0.054, 0.09) determined by measuring the Zn, Mg and Te contents with accuracy better than 1% applying the inductively coupled plasma-Optical Emission Spectrometry (ICP-OES) method[17] to powders. All samples crystallize in the same zincblende structure as ZnTe (Fig. 1a), verified by powder X-ray diffraction either in laboratory conditions at ambient pressure using the Cu K$\alpha$ line at 1.541 Å or on the PSICHÉ beamline of synchrotron SOLEIL at nearly ambient pressure ($x$=0.046, 0.38 GPa) using the 0.3738 Å radiation. The lattice constant exhibits a linear composition dependence (dashed curve in the inset of Fig. 1a) between the ZnTe (open symbol)[18] and MgTe[19] values, as expected in case of a regular Zn↔Mg substitution on the cationic site.[20]

The nature of the Zn↔Mg substitution, as to whether this is ideally random or due to clustering or anticlustering is studied by powder $^{125}$Te solid-state nuclear magnetic resonance (NMR). Technical details is given as supplementary material. The $^{125}$Te chemical shift is sensitive to the electronic environment and changes with the first-neighbor tetrahedral environment (from 0 to 4 Zn/Mg) of the Te nucleus. The relative abundance of the various Te-centered tetrahedra forming $Zn_{0.91}Mg_{0.09}Te$ can be derived from the integrals of the $^{125}$Te signals, presently modeled as Gaussian functions (curve, Fig. 1b), using the dmfit program.[21] $Zn_{0.91}Mg_{0.09}Te$ appears to be formed of three elementary Te-centered tetrahedra with 4,3 or 2 Zn atoms at the vertices; the Mg-rich tetrahedra are absent. The measured tetrahedra fractions by NMR (Fig. 1b), *i.e.*, ($f_4$=0.66, $f_3$=0.28, $f_2$=0.06), give a perfect match (within experimental resolution, $\Delta f$=±0.02%) with predictions by the Bernouilli's binomial distribution ($f_4$=0.68, $f_3$=0.27, $f_2$=0.04). The Zn↔Mg substitution is thus random in $Zn_{0.91}Mg_{0.09}Te$, and presumably also in all studied crystals given their close Mg contents. Altogether, the $Zn_{1-x}Mg_xTe$ lattice structure looks ideal at both the macroscopic (X-ray diffraction) and microscopic (NMR) scales, as needed to achieve a reliable insight into the lattice dynamics.

The theoretical $q$-dependent $Zn_{0.91}Mg_{0.09}Te$ PP Raman cross section outlined in Fig. 2, informing on the PP dispersion (Raman frequency – clear/bright plain curves) and on the PP Raman intensity (thickness/brightness of curves), provides an overview of all accessible PP by Raman scattering depending on the used backward ($q \to \infty$) or forward ($q \to 0$) scattering geometry. Similar information for pristine ZnTe is given as supplementary material (prefixed S, Fig. S1), for reference purpose.

The $Zn_{0.91}Mg_{0.09}Te$ (resp. ZnTe) PP dispersions are governed by two sets of asymptotes.[1-3] On the one hand, four (resp. two) horizontal phonon asymptotes are pinned to the ($\omega_T$, $\omega_L$) frequencies of the Zn-Te and Mg-Te PM-TO ($q \to \infty$) and longitudinal optical (LO, $q \to 0$) modes. On the other hand, there are two quasi vertical photon asymptotes (solid lines) governed by the static ($\varepsilon_s$, $\omega \ll \omega_T$) and high-frequency ($\varepsilon_\infty$, $\omega \gg \omega_T$) relative dielectric constants of the crystal. The strong PP-coupling occurs at the crossing of the phonon and photon asymptotes. In ZnTe (Fig. S1), this gives rise to two $PP^\pm$ (in order of frequency) coupled modes in mutual repulsion with mixed phonon (dominant at $q \to \infty$ for $PP^-$ and at $q \to 0$ for $PP^+$) and photon (dominant at $\omega \to 0$ for $PP^-$ and at $\omega \to \infty$ for $PP^+$) characters. The Raman intensity is large when the PP is phonon-like and vanishes to zero when it turns photon-like (Raman extinction – Ext.). $Zn_{1-x}Mg_xTe$-alloying introduces an intermediary $PP^{int}$ branch between the hyperbolic $PP^\pm$ ones. This exhibits a $S$-like dispersion marked by a Raman extinction near the inflexion (Fig. 2).[6,22] Strictly at $q$=0 the PP (TO) and LO modes are degenerate.[23] The latter are dispersionless, vibrating at higher frequency than the former ($\omega_L > \omega_T$)[24] in the pristine crystals.[25]

The $q$-dependent $Zn_{0.91}Mg_{0.09}Te$ PP Raman cross section (Fig. 2) has been calculated via the generic formula set in Ref.[5] (specifically using Eq. 1 therein) assuming that the amounts of phonon oscillator strength $OS = \varepsilon_\infty \times \omega_T^{-2} \times (\omega_L^2 - \omega_T^2)$[12] and Faust-Henry coefficient $C_{FH}$[26] (that measures the relative Raman scattering efficiencies by the atom displacement and by the related electric field of a lattice vibration) awarded to the Zn-Te and Mg-Te modes scale as the corresponding bond fractions. The used ($\varepsilon_\infty$[7,8,9], $\omega_T$, $\omega_L$, $C_{FH}$) values are (7.2, 176 cm$^{-1}$, 205 cm$^{-1}$, -0.11)[5] for ZnTe and (5.72,[27] 235 cm$^{-1}$,[9] 292 cm$^{-1}$,[9] -0.17) for MgTe. $C_{FH}$ for MgTe, not documented in the literature, is adjusted with accuracy

±0.02, so that the theoretical ratio between the intensities of the Zn-Te and Mg-Te signals achieved in the LO version of the $Zn_{0.91}Mg_{0.09}Te$ Raman cross section[5] matches experiment (see below). Two additional alloy-related input parameters, *i.e.*, the Zn-Te (172 cm$^{-1}$) and Mg-Te (255 cm$^{-1}$) PM-TO frequencies at $x$=0.09, are close to the reported values at $x$=0.1 in the literature.[11] The Zn-Te frequency is observed in the (TO-allowed, LO-forbidden) Raman spectra taken in backscattering on a cleaved (110) face of $Zn_{0.91}Mg_{0.09}Te$ (Fig. 3a – thick curve).[28] The minor PM-TO Mg-Te mode is hardly visible. First, because it is weak at small Mg content. Next, because it is obscured by second-order (symbolized ×2) modes.[9,10] Finally, because various stable Mg isotopes provide slightly distinct signals,[9] adding to the confusion. The Mg-Te Raman signal clarifies only from $x$=0.4 onwards.[9-11] At 77 K, the second-order signal collapses, which "purifies" the first-order one. In the Mg-Te range, the small peak at ~260 cm$^{-1}$ grows with the Mg content (Fig. 3b), hence assigned as the searched PM-TO Mg-Te mode at 77 K. *Ab initio* calculations at 0 K of the local vibration modes due to an isolated Mg-duo[29] (for three stable Mg isotopes) in ZnTe predict the Mg-Te PM-TO mode in the range 255 – 266 cm$^{-1}$ (Fig. S2). The *ab initio* (255 – 266 cm$^{-1}$, 0 K) and Raman (~260 cm$^{-1}$, 77 K) estimates are consistent with the PM-TO Mg-Te frequency (255 cm$^{-1}$) used to generate the PP Raman cross section at 300 K (Fig. 2), taking into account that the Raman frequency increases by reducing the temperature, due to the shrinking of the lattice, that enlarges the bond force constants.

A selection of near-forward Raman spectra taken through parallel naturally oriented (110) $Zn_{0.91}Mg_{0.09}Te$ crystal faces obtained by cleavage at various scattering angles $\theta$ between the incident-$\vec{k}_i$ and scattered-$\vec{k}_s$ wavevectors inside the crystal using the 632.8 nm laser line is shown in Fig. 3a (top panel). While theoretically forbidden, the LO modes are activated by multi-reflection of the laser beam between the top and bottom crystal faces.[3] This spurious LO signal is used to adjust $C_{FH}$ for MgTe (not shown), as discussed above. Both the PM-TO and PP variants of the allowed Zn-Te and Mg-Te TO modes are visible. This is because the nominal/dominant PP signal generated by the laser beam on its way forth towards the top crystal face superimposes onto the spurious/minor PM-TO signal produced by the laser beam on its way back to the bottom crystal face after reflection at the top crystal face.

The proper $\theta$ value per PP spectrum is estimated by tuning the $\omega(q)$ Raman scan line (dashed curves) for the used scattering geometry until this scan line crosses the $Zn_{0.91}Mg_{0.09}Te$ PP dispersion (Fig. 2) exactly at the experimentally observed $PP^-$ and $PP^{int}$ frequencies. Technically, the $\omega(q)$ Raman scan line is derived from the wavevector conservation rule ($\vec{q} = \vec{k}_i - \vec{k}_s$) that governs the Raman scattering, expressing $k_{i,s}$ as $n(\omega_{i,s}) \times \omega_{i,s} \times c^{-1}$, where $c$ is the speed of light in vacuum, $\omega_{i,s}$ the frequency of the incident/scattered light, and $n(\omega_{i,s})$ the related refractive index of the crystal.[13] In the used Stokes geometry $\omega = \omega_i - \omega_s$. The $\theta$-dependence occurs via $\vec{k}_i \cdot \vec{k}_s$, scaling as $\cos\theta$.

Various $\theta$ angles are probed in Fig. 3a. At any $\theta$ value, $PP^{int}$ emerges next to its native Mg-Te PM-TO mode. The Raman intensity is large or small depending on whether the Raman scan line intercepts the $PP^{int}$ dispersion beyond (small $\theta$ value) or above (large $\theta$ value) the Raman extinction (Ext., Fig. 2), respectively. $PP^-$ is more $\theta$-sensitive. At large angle ($\theta$~2.4°) $PP^-$ is probed at the onset of the PP-regime where it remains closely attached to its native Zn-Te PM-TO mode prior to suffering a transient photon-like Raman extinction (Ext.) at the entry of the bottleneck of the $PP^-$ dispersion. Small angles ($\theta$~0) are achieved beyond the Raman extinction, marking a reactivation of the $PP^-$ Raman signal at a much lower frequency than the native Zn-Te PM-TO mode (by as much as ~35 cm$^{-1}$), as predicted (Fig. 2). $PP^-$ is dramatically enhanced when $\theta$ achieves minimum. Regardless of the $PP^-$ enhancement, even the reactivation of $PP^-$ beyond the extinction is hardly observed with $Zn_{0.954}Mg_{0.046}Te$ (Fig. S3), and just absent with ZnTe (Fig. S3 of Ref.[5]). Such discrepancy cannot be anticipated since the theoretical $PP^-$ dispersion and Raman intensity are so similar for ZnTe, $Zn_{0.954}Mg_{0.046}Te$ and $Zn_{0.91}Mg_{0.09}Te$ (compare, *e.g.*, Figs. 2 and S1). The $PP^-$ enhancement in the photon-like regime is thus a positive surprise due to the alloy disorder.

The enhanced $PP^-$ mode seen with $Zn_{0.91}Mg_{0.09}Te$ by using the 632.8 nm laser excitation exhibits many interesting features. **(i)** It is massively supported by the crystal, testified by an intense Raman signal. **(ii)** It "feels" like propagating in a pristine crystal, being "blind" to the chemical disorder and to the local lattice distortions inherent to alloying. Indeed, its Raman intensity obeys the same nominal

Raman selection rules as the PM-TO mode of pure ZnTe (Fig. S4 of Ref.[5]) when rotating as a solid body over 360° at a given spot of the crystal surface the polarizations of the incident laser and of the scattered light taken parallel ($\vec{e}_i \parallel \vec{e}_s$, Fig. 3c). **(iii)** Being probed in the bottleneck of the $PP^-$ dispersion at the onset of the steep photon-like dispersion, it propagates at a high (photon-like) velocity. **(iv)** It exhibits a long lifetime, reflected by a sharp Raman signal (full width at half maximum ~13 cm$^{-1}$, slightly less than twice that of the native Zn-Te PM-TO mode, *i.e.*, ~7 cm$^{-1}$).

By using the 785 nm laser line, $PP^-$ is probed deeper downward the asymptotic photon-like regime. Consider an ideal forward scattering geometry ($\theta=0°$). In this case, $q$ achieves minimum, *i.e.*, $q_{min} = c^{-1} \times |n(\omega_i) \times \omega_i - n(\omega_s) \times \omega_s|$. In the used Stokes geometry ($\omega_i > \omega_s$) $q_{min}$ reduces by changing the laser line from 632.8 nm to 785 nm because the refractive index of Zn$_{1-x}$Mg$_x$Te is less dispersive around 785 nm than around 632.8 nm, at any composition $x$.[13] When $\theta$ achieves minimum, the attractive $PP^-$ features (i) intensity (Fig. 3a – bottom panel), (ii) selection rules (Fig. S4c) and (iii) velocity (Fig. 2), are preserved. However, $PP^-$ is overdamped (Fig. 3a – bottom panel), so that feature (iv) small width is missing. This is due to the finite angles of the incidence ($\vec{k}_i$) and detection ($\vec{k}_s$) cones in a near-forward Raman scattering experiment. Due to this experimental limitation a finite $\theta$-domain is probed, and not a unique $\theta$ value. The $PP^-$ frequency domain is correspondingly large, since the addressed photon-like $PP^-$ dispersion at $\theta$~0° by using the 785 nm laser line is so steep (Fig. 2).

Summarizing, the phonon-photon coupled mode (phonon-polariton) probed at minor Mg-alloying of ZnTe by using the 632.8 nm laser excitation seems optimal. It exhibits a long lifetime, is strongly bound to the lattice, yet remains insensitive to the alloy disorder and propagates at a high velocity. Such mode is absent of ZnTe. Hence, alloying proves beneficial in view to achieve ultrafast (photon-like) signal processing in matter (phonon-like) in the terahertz range of lattice vibrations – with potential applications in photonics, beyond what is possible with the pristine crystals.

## SUPPLEMENTARY MATERIAL

Supplementary material serves several purposes. First, detail is given concerning the used experimental setup to perform the solid-state NMR measurements. Second, the $q$-dependent phonon-polariton Raman cross section of ZnTe is introduced (Fig. S1) for comparison with that of $Zn_{1-x}Mg_xTe$ at minor Mg content. Third, *ab initio* calculations of the local vibrational modes due to various Mg-duos made of stable Mg isotopes immersed in ZnTe (Fig. S2) are performed in view to guide experimental studies of the Mg-Te impurity mode/band at minor Mg-alloying of ZnTe (Fig. 3a). Fourth, we report on a near-forward Raman study of $Zn_{0.955}Mg_{0.045}Te$ using the 632.8 nm laser excitation (Fig. S3), for completeness. Last, the raw polarized near-forward $Zn_{0.91}Mg_{0.09}Te$ Raman spectra used to generate the butterfly pattern displayed in Fig. 3c (symbols) are made available (Fig. S4)

## ACKNOWLEDGEMENTS

We acknowledge assistance from the PSICHÉ beamline staff of synchrotron SOLEIL (Proposal 20210410 – BAG for PSICHÉ beamline led by Y.L.G., sub-project HP-ZnMgTe co-led by A.P. & O.P.), from the IJL core facility (Université de Lorraine – http://ijl.univ-lorraine.fr/recherche/centres-de-competences/rayons-x-et-spectroscopie-moessbauer-x-gamma) for the X-ray diffraction measurements at ambient pressure. VJBT in charge with the *ab initio* AIMPRO calculations C.acknowledge the FCT through projects LA/P/0037/2020, UIDB/50025/2020 and UIDP/50025/2020. *This work was supported by the French PIA project Lorraine Université d'Excellence, part of the France 2030 Program, reference ANR-15-IDEX-04-LUE, within* the ViSA – IRP *«***Vi***brations of* **S***emiconductor* **A***lloys –* **I***nternational* **R***esearch* **P***artnership»* wall-less associated international laboratory (2024 – 2028), co-funded by the Excellence Initiative – Research University program at Nicolaus Copernicus University in Toruń.

## AUTHOR DECLARATIONS

### Conflict of Interest

The authors have no conflicts to disclose.

## AUTHOR CONTRIBUTIONS

**Diksha Singh:** Formal analysis (equal); Investigation (equal). **Abdelmajid Elmahjoubi:** Formal analysis (lead); Investigation (equal); Software (equal); Validation (equal); Visualization (lead). **Olivier Pagès:** Conceptualization (lead); Funding acquisition (lead); Methodology (lead); Project administration (lead); Supervision (lead); Visualization (equal); Writing – original draft preparation (lead). **Vitor J. B. Torres:** Formal analysis (equal); Investigation (equal); resources (equal); Software (equal). **Carole Gardiennet:** Formal analysis (equal); Investigation (equal). **Gwendal Kervern:** Formal analysis (equal); Investigation (equal). **Alain Polian:** Investigation (equal); Project administration (equal); validation (equal). **Yann Le Godec:** Investigation (equal); project administration (equal); Resources (equal). **Jean-Paul Itié:** Investigation (equal); project administration (equal); Resources (equal). **Sébastien Diliberto:** Formal analysis (equal); Investigation (equal). **Stéphanie Michel:** Formal analysis (equal); Investigation (equal). **Pascal Franchetti:** Resources (equal). **Karol Strzałkowski:** Project administration (equal); Resources (equal); Supervision (equal).

## DATA AVAILABILITY

The data that support the findings of this study are available from the corresponding author upon reasonable request.

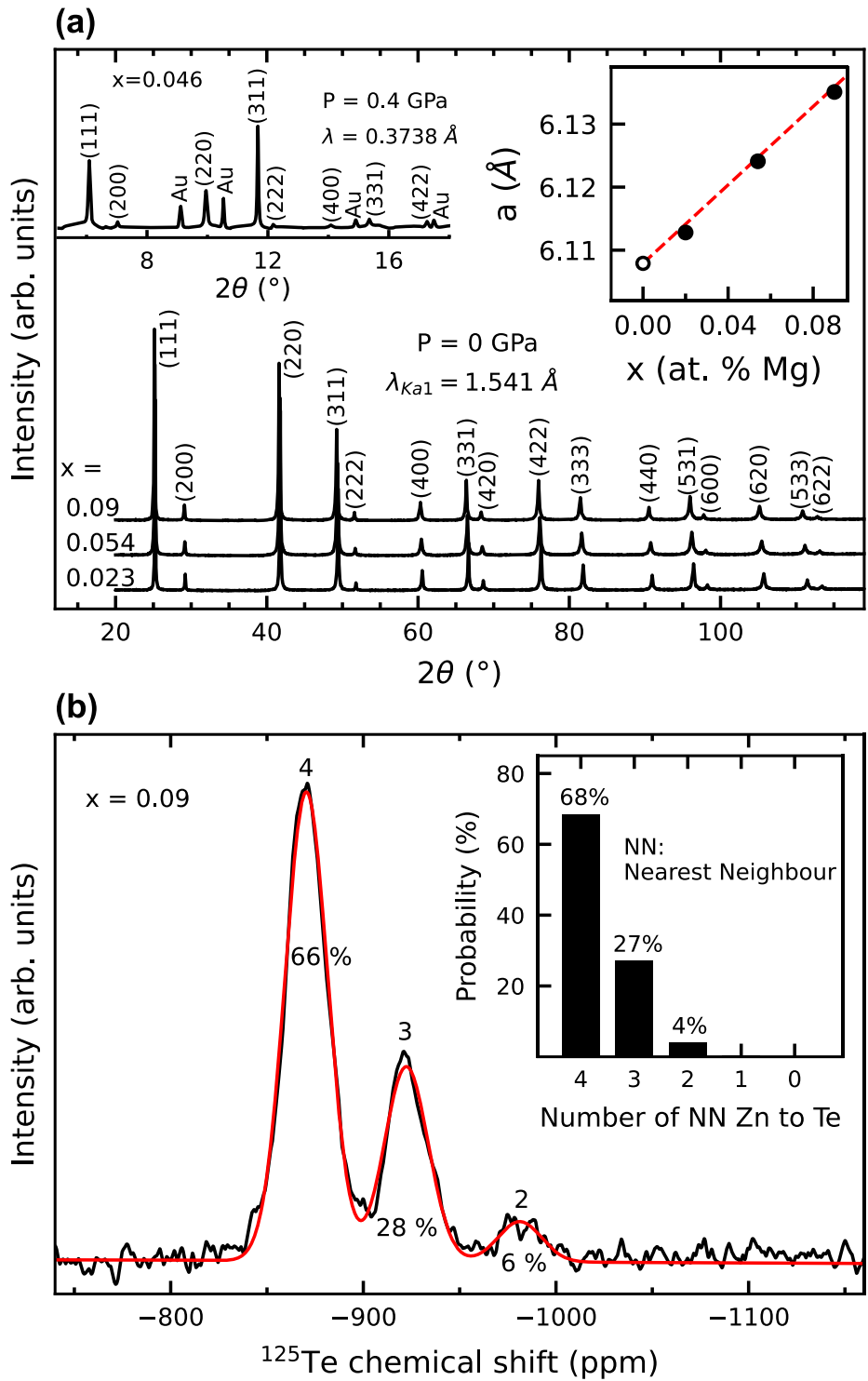

**FIG. 1. Zn$_{1-x}$Mg$_x$Te ($x \leq 0.09$) lattice structure.** (a) Powder Zn$_{1-x}$Mg$_x$Te X-ray diffractograms obtained at ambient pressure in laboratory and at nearly ambient pressure (as specified) on the PSICHÉ beamline of synchrotron SOLEIL (left inset). Diffraction lines due to Au used for pressure calibration are indicated. Right inset: Linear composition dependence of the lattice constant at 0 GPa (dashed curve) between the ZnTe (hollow symbol – taken from Ref. 18) and MgTe (taken from Ref. 19) values. Experimental data are superimposed (solid symbols) (b) Powder Zn$_{0.91}$Mg$_{0.09}$Te $^{125}$Te solid-state NMR spectra. The NMR peaks are modeled by using three Gaussian functions (curve). Inset: Bernoulli binomial distribution of Te-centered tetrahedra depending on the number of nearest-neighbor (NN) Zn atoms at the vertices achieved in case of a random Zn↔Mg substitution.

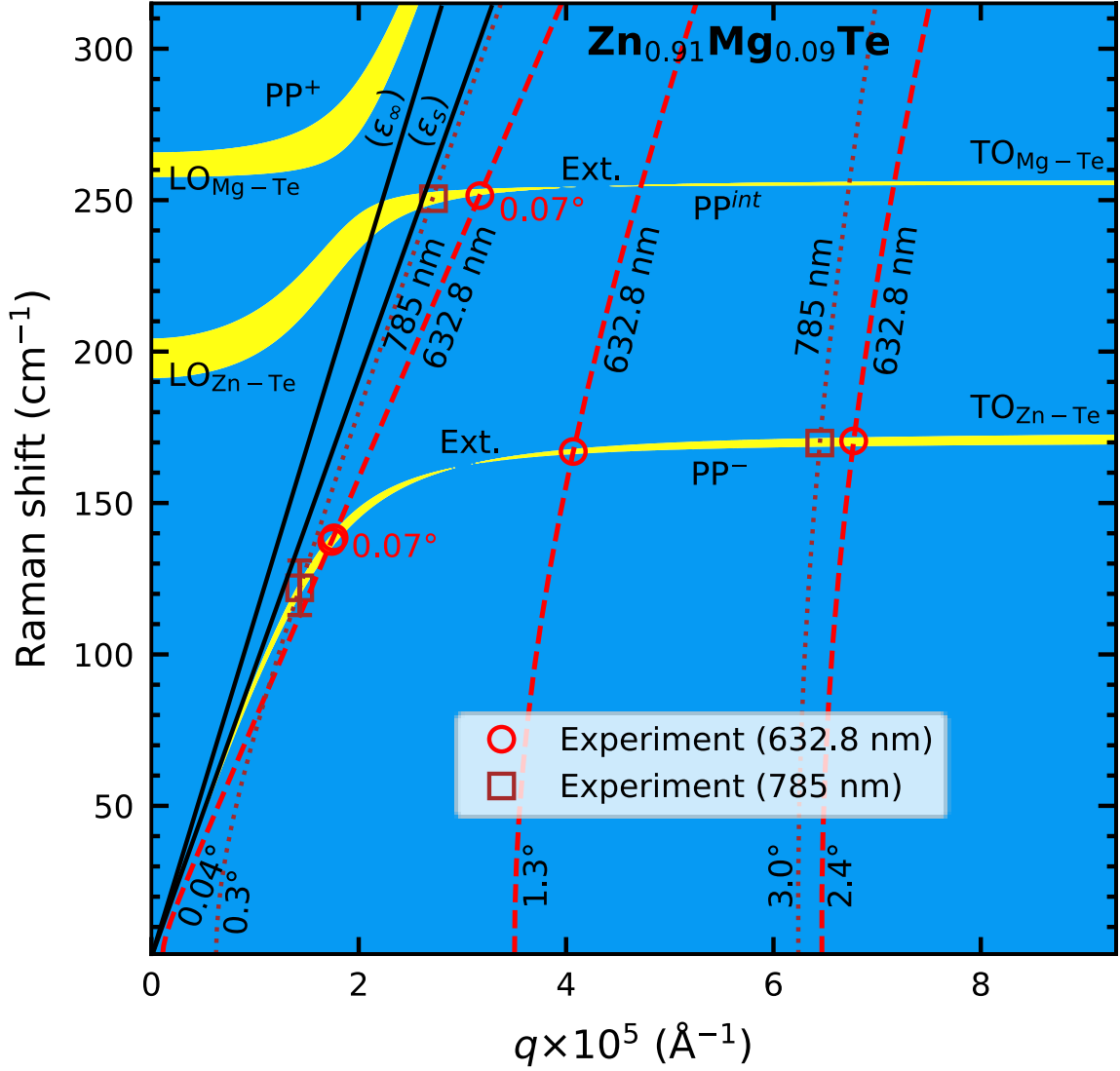

**FIG. 2. Zn$_{0.91}$Mg$_{0.09}$Te lattice dynamics.** PP dispersion (bright curves) and PP Raman intensity (thickness/brightness of curves) generated by the $q$-dependent Zn$_{0.91}$Mg$_{0.09}$Te PP Raman cross section. The Raman scan lines for various scattering angles ($\theta$) addressed in Fig. 3 using the 785 nm (dotted curves) and 632.8 nm (dashed curves) laser excitations reveal a series of PP modes (symbols, the error bar is shown when exceeding the symbol size). In absence of Raman scan line, the $\theta$ value is specified next to the symbol. $\varepsilon_S$ and $\varepsilon_\infty$ govern the photon asymptotes at low- and high-frequency (straight plain curves), respectively.

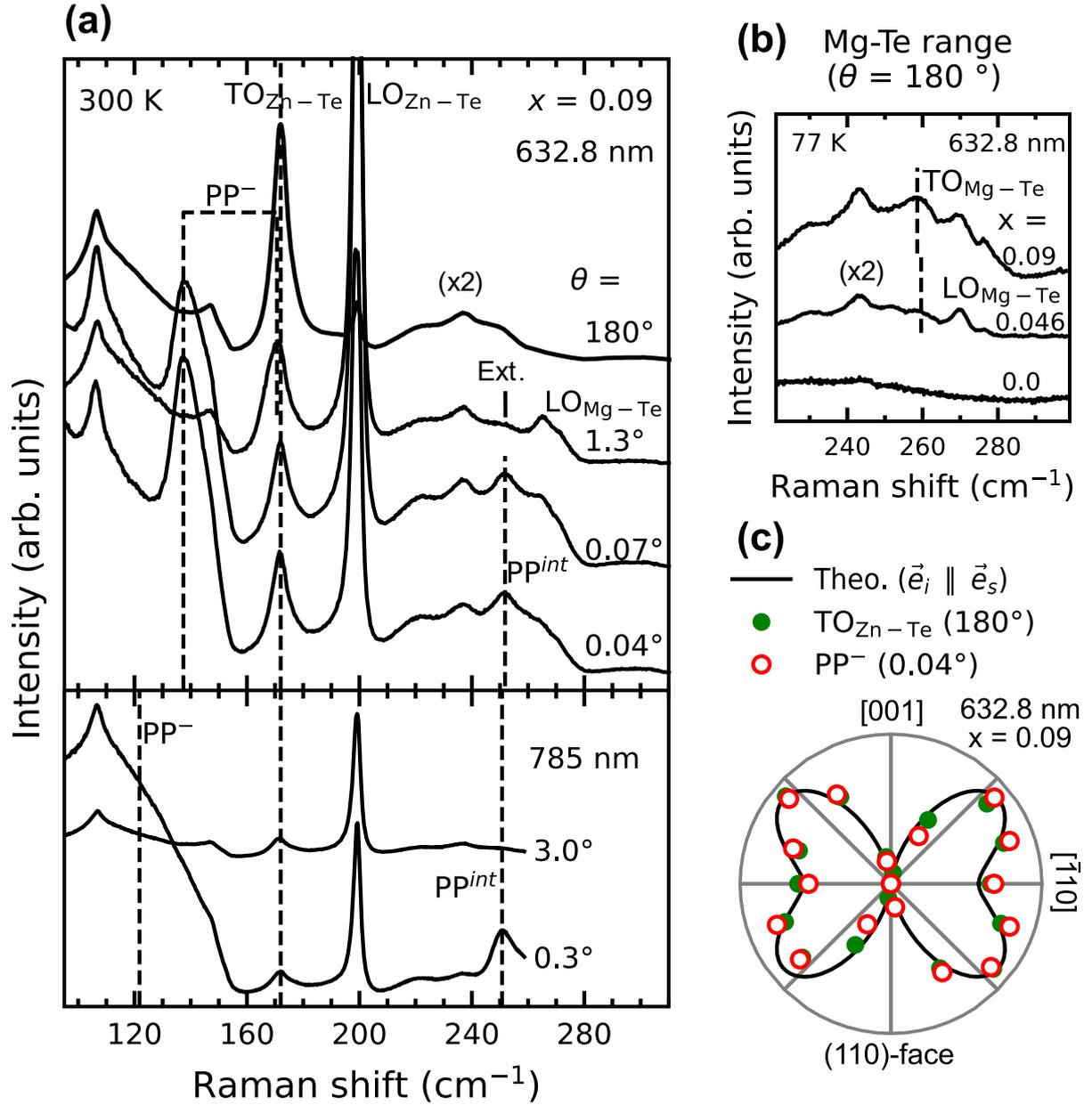

**FIG. 3. Zn$_{0.91}$Mg$_{0.09}$Te $\theta$-dependent Raman spectra.** (a) Backward (thick curve) and near-forward (thin curves) Zn$_{0.91}$Mg$_{0.09}$Te Raman spectra at 300 K taken with the 632.8 nm (upper panel) and 785 nm (lower panel) laser excitations at various scattering angles ($\theta$), as indicated. The $\theta$-dependence of PP up to the native TO modes is emphasized (dashed curves). (b) Mg-Te Raman signal at 77 K. (c) Near-forward $PP^-$ ($\theta=0.04°$, hollow symbols) and backward $TO_{Zn-Te}$ (filled symbols) Raman selection rules measured on cleaved (110) crystal faces at the same sample spot in parallel polarizations ($\vec{e}_i$, $\vec{e}_s$) with the 632.8 nm laser excitation.

**Supplementary Material for "Raman scattering by fast lattice-supported phonon-polaritons in Zn$_{1-x}$Mg$_x$Te alloys ($x \leq 0.09$)"**

D. Singh,[1,2] A. Elmahjoubi,[1] O. Pagès,[1,a)] V.J.B. Torres,[3] C. Gardiennet,[4] G. Kervern,[4] A. Polian,[5] Y. Le Godec,[5] J.-P. Itié,[6] S. Diliberto,[7] S. Michel,[7] P. Franchetti,[1] and K. Strzałkowski[2]

1. Université de Lorraine, LCP-A2MC, UR 201019679B, F-57000 Metz, France
2. Institute of Physics, Faculty of Physics, Astronomy and Informatics, Nicolaus Copernicus University in Toruń, ul. Grudziądzka 5, 87-100 Toruń, Poland
3. Departamento de Fisica and I3N, Universidade de Aveiro, 3810 – 193 Aveiro, Portugal Institut de
4. Université de Lorraine, Laboratoire de Cristallographie, Résonance Magnétique et Modélisations, UMR 7036, Vandoeuvre-lès-Nancy, F-54506, France
5. Institut de Minéralogie, de Physique des Matériaux et de Cosmochimie, Sorbonne Université — UMR CNRS 7590, F-75005 Paris, France
6. Synchrotron SOLEIL, L'Orme des Merisiers Saint-Aubin, BP 48 F-91192 Gif-sur-Yvette Cedex, France
7. Université de Lorraine, CNRS, IJL, F-57000, Metz, France

a) Author to whom correspondance should be addressed : olivier.pages@univ-lorraine.fr

The following deals with various side issues related to the discussion done in the main text. In **Sec. SI**, detail is given concerning the experimental conditions used to record the reported Zn$_{0.91}$Mg$_{0.09}$Te NMR data (Fig. 1b). **Sec. SII** reports on *ab initio* calculations of the Zn$_{1-x}$Mg$_x$Te Raman spectra in the Mg-dilute limit ($x \sim 0$) for various stable Mg isotopes. This helps to identify the native Mg-Te phonons behind the MgTe-like PP, that hardly shows up in the reported Zn$_{1-x}$Mg$_x$Te ($x \leq 0.09$) Raman spectra at 300 K. **Sec. SIII** provides the raw polarized near-forward Zn$_{0.91}$Mg$_{0.09}$Te Raman spectra behind the reported Raman selection rules for the enhanced-$PP^-$ features observed with the 632.8 nm laser excitation at minimal scattering angle (Fig. 3c). A similar insight for the overdamped-$PP^-$ feature probed with the 785 nm laser excitation is also provided, for completeness.

Before entering the discussion of the lattice structure and of the PP-related lattice dynamics of Zn$_{1-x}$Mg$_x$Te at minor Mg-alloying, we find it useful to introduce in Fig. S1 the ZnTe PP Raman cross section calculated in its $q$-dependence using the set input parameters ($\varepsilon_\infty, \omega_T, \omega_L, C_{FH}$) in the main text, for reference purpose. The Raman scan lines in the perfect forward scattering geometry ($\theta=0°$) related to the 785 nm and 632.8 nm laser excitations are superimposed to appreciate which limit PP Raman modes can be achieved experimentally using such excitations.

I. Zn$_{0.91}$Mg$_{0.09}$Te: Solid-state nuclear magnetic resonance measurements

The $^{125}$Te solid-state nuclear magnetic resonance (NMR) data reported in Fig. 1b provide a direct evidence that the Zn↔Mg substitution is ideally random in the studied Zn$_{0.91}$Mg$_{0.09}$Te crystal.

$^{125}$Te NMR spectra were acquired on a Bruker Avance III 600 MHz spectrometer (14.1 T) equipped with a Bruker 2.5 mm double resonance MAS probe. The X channel was tuned to $^{125}$Te frequency (189.3 MHz). The $^1$H channel was not used, as no proton decoupling was necessary. $^{125}$Te direct acquisition spectra were recorded at 25 kHz spinning frequency using 3.7 μs $^{125}$Te 90° pulses. 40 transients were collected with a 10000 s interscan delay, corresponding to at least 5 times the longest measured longitudinal relaxation time T$_1$ (saturation recovery experiment, not shown). The NMR spectra were processed by using a Gaussian apodization function with a 400 Hz line broadening and signals were deconvoluted using a gaussian function. $^{125}$Te chemical shift was externally calibrated using a ZnTe powder sample (containing Te-clusters of the 0-Mg type only) resonating at -888 ppm, consistently with existing measurements in the literature.[S1]

II. Zn$_{1-x}$Mg$_x$Te ($x$~0): *Ab initio* Raman spectra

To assign the minor Mg-Te Mg in ZnTe ($x$~0), hardly visible in the reported Zn$_{1-x}$Mg$_x$Te Raman spectra ($x$ ≤0.09) at 300 K, we perform *ab initio* calculations using the AIMPRO code[14-15] of the Raman spectrum related to a large (216-atom) ZnTe zincblende-type supercell containing an isolated Mg-duo (Fig. S2) marking the onset of significant Mg-alloying at $x$ ≤0.09 (refer to the NMR data – Fig. 1). In its current version the AIMPRO code does not handle the macroscopic ($q$~0) electric field accompanying a polar (LO or PP) Raman mode, and thus provides an insight limited to the PM-TO modes, of main interest for our use (referring to the input parameters of the Raman cross section – see main text).

The *ab initio* calculations are done as described in Ref.[S2] The linear density approximation (LDA) is used for the exchange-correlation potential by Perdew and Wang,[S3] taking the norm-conserving pseudopotentials for Zn, Mg and Te as proposed by Hartwigsen *et al*.[S4] The 216-atom cubic supercells are sampled by a 2×2×2 special k-point mesh as proposed by Monkhorst and Pack,[S5] using a third-order Birch-Murnaghan[S6] equation of state. The phonon calculations are done after full relaxation (atom positions and lattice constant) of the supercells.

This approach tested on a reference zincblende-type ZnTe (216-atom) supercell gives the PM-TO mode at $\omega_{T,Zn-Te}$=191.9 cm$^{-1}$, slightly above that found experimentally by Raman scattering with ZnTe by Camacho *et al*. (177.5 cm$^{-1}$)[S7] and in this work (176 cm$^{-1}$). This is due to a well-know bias of the LDA to underestimate the lattice constant, and thus to overestimate the bond force constants.

The $^{24}$Mg-duo generates six (three degrees of freedom per atom) local vibrational modes regrouped into three distinct peaks in the range 263 – 277 cm$^{-1}$, downshifted by 4-5 cm$^{-1}$ with the $^{25}$Mg-duo and again by the same amount with $^{26}$Mg-duo. Altogether, the current *ab initio* data predict the PM-TO Mg-Te Raman mode at minor Mg-alloying of ZnTe in the range 255 – 266 cm$^{-1}$. This is consistent with the retained value of 256 cm$^{-1}$ for the Mg-Te PM-TO frequency in the calculation of the Zn$_{0.91}$Mg$_{0.09}$Te Raman cross section (Fig. 2).

III. Zn$_{1-x}$Mg$_x$Te experimental Raman spectra

1. Zn$_{0.955}$Mg$_{0.45}$Te: near-forward Raman study ($x$ ≤0.09)

Fig. S3 displays a selection of unpolarized near-forward Raman spectra obtained with Zn$_{0.955}$Mg$_{0.045}$Te using the 632.8 nm laser excitation at representative scattering angles ($\theta$). The proper $\theta$ value per spectrum is estimated by adjusting the $\theta$-dependent Raman scan line until it crosses the $q$-dependent PP Raman cross section of Zn$_{0.955}$Mg$_{0.45}$Te (similar in every respect to that reported for Zn$_{0.91}$Mg$_{0.09}$Te in Fig. 2, hence not shown) exactly at the experimentally observed $PP^-$ and $PP^{int}$ frequencies (as explained in the main text). By decreasing $\theta$, the $PP^-$ and $PP^{int}$ modes progressively depart from their native Zn-Te and Mg-Te PM-TO modes ($\theta$~2.4°), respectively. The $PP^-$ feature, of major interest, suffers a Raman extinction at the entry of the highly-dispersive photon-like regime and is reactivated at minimal scattering angle ($\theta$~0°), as observed with Zn$_{0.91}$Mg$_{0.09}$Te (Fig. 3 – top panel). However, the reactivated $PP^-$ mode of Zn$_{0.955}$Mg$_{0.045}$Te fails to develop into a strong feature, in contrast with that observed with Zn$_{0.91}$Mg$_{0.09}$Te.

2. Zn$_{0.91}$Mg$_{0.09}$Te: $PP^-$ near-forward Raman selection rules ($\theta$~0)

The raw near-forward Raman spectra behind the Raman selection rules (Fig. 3c – symbols) obeyed by the enhanced-$PP^-$ mode of Zn$_{0.91}$Mg$_{0.09}$Te in parallel ($\vec{e}_i \parallel \vec{e}_s$) polarizations at quasi normal incidence/detection ($\theta$~0, $\vec{k}_i \parallel \vec{k}_s$) across cleaved (110) crystal faces using the 632.8 nm laser excitation are shown *in extenso* in Fig. S4b. Corresponding Raman spectra related to the overdamped-$PP^-$ mode and to the native Zn-Te PM-TO mode (behind $PP^-$) of Zn$_{0.91}$Mg$_{0.09}$Te observed at the same sample spot with the 785 nm and 632.8 nm laser excitations, respectively, are displayed in Figs. S4c and S4a, respectively. The three series of spectra are labeled according to the angle $\alpha$ between the incident polarization ($\vec{e}_i$) and the [$\bar{1}$10] in-plane crystal axis, taken as an arbitrary reference.

The enhanced-$PP^-$ (632.8 nm), overdamped-$PP^-$ (785 nm) and PM-TO signals exhibit similar $\alpha$-dependencies. In fact, the enhanced-$PP^-$ experimental data (Fig. 3c, symbols) ideally reproduce the same theoretical butterfly-like pattern (curves) as the native Zn-Te PM-TO mode (behind $PP^-$) of Zn$_{0.91}$Mg$_{0.09}$Te. The three modes are thus identical in nature, as many variants of the same Zn-Te TO mode, either probed in its polar (overdamped/enhanced-$PP^-$) or non-polar (PM-TO) regimes.

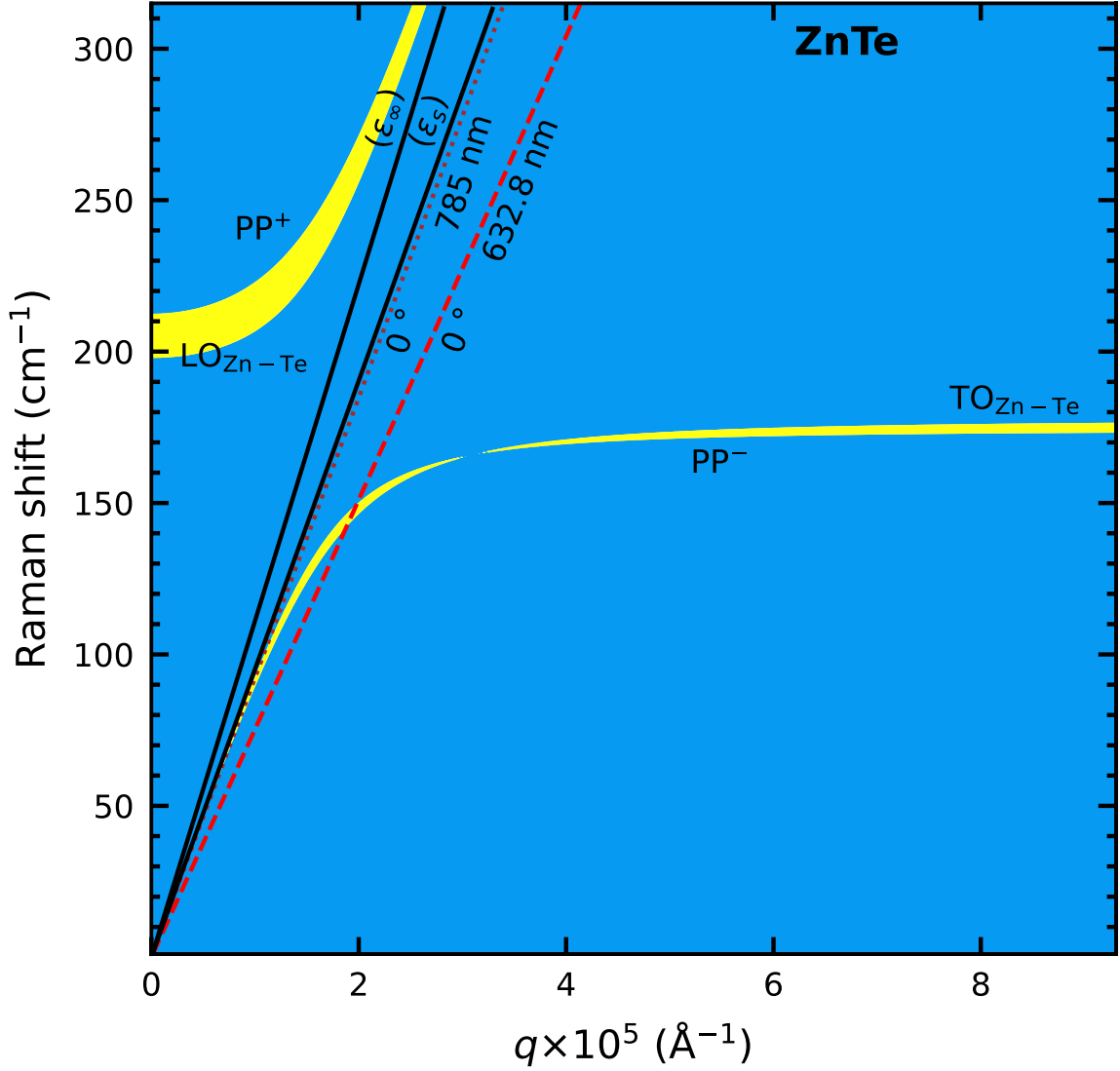

**FIG. S1. ZnTe $q$-dependent phonon-polariton Raman cross section.** PP dispersion (bright plain curves) and PP Raman intensity (thickness/brightness of curves) calculated from the $q$-dependent ZnTe PP Raman cross section. The Raman scan lines at zero scattering angle ($\theta$) for the 785 nm (dotted curves) and 632.8 nm (dashed curves) laser excitations are shown. The photon asymptotes at low- and high-frequency (straight plain curves), governed by $\varepsilon_S$ and $\varepsilon_\infty$, respectively, are indicated.

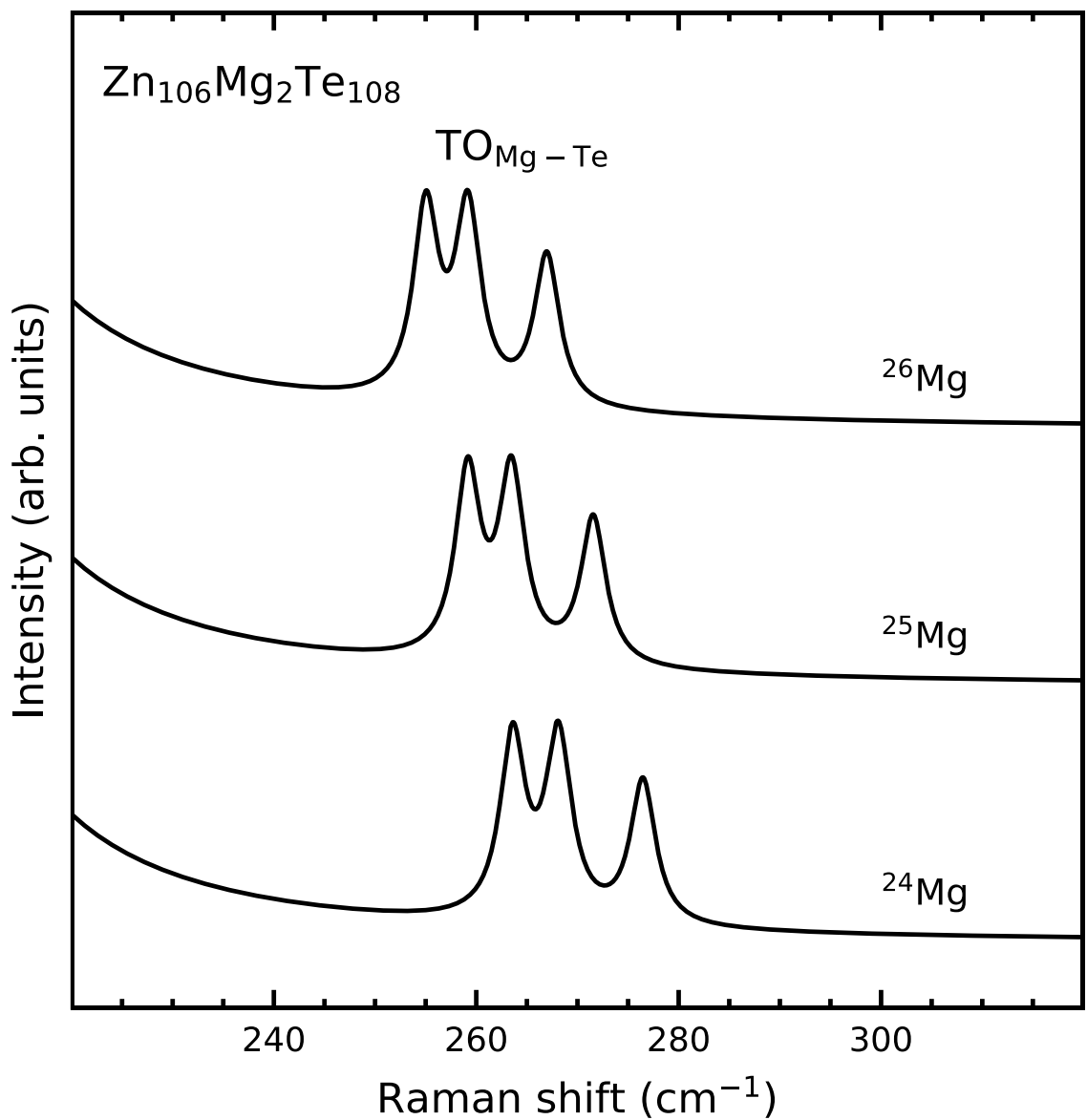

**FIG. S2. *Ab initio* Mg-Te Raman signal of Zn$_{1-x}$Mg$_x$Te ($x$~0).** Local vibrational modes (three par atom, partially degenerated) calculated *ab initio* (AIMPRO) for three Mg-duos due to various stable Mg isotopes (as indicated) isolated in large ZnTe (216-atom) supercells.

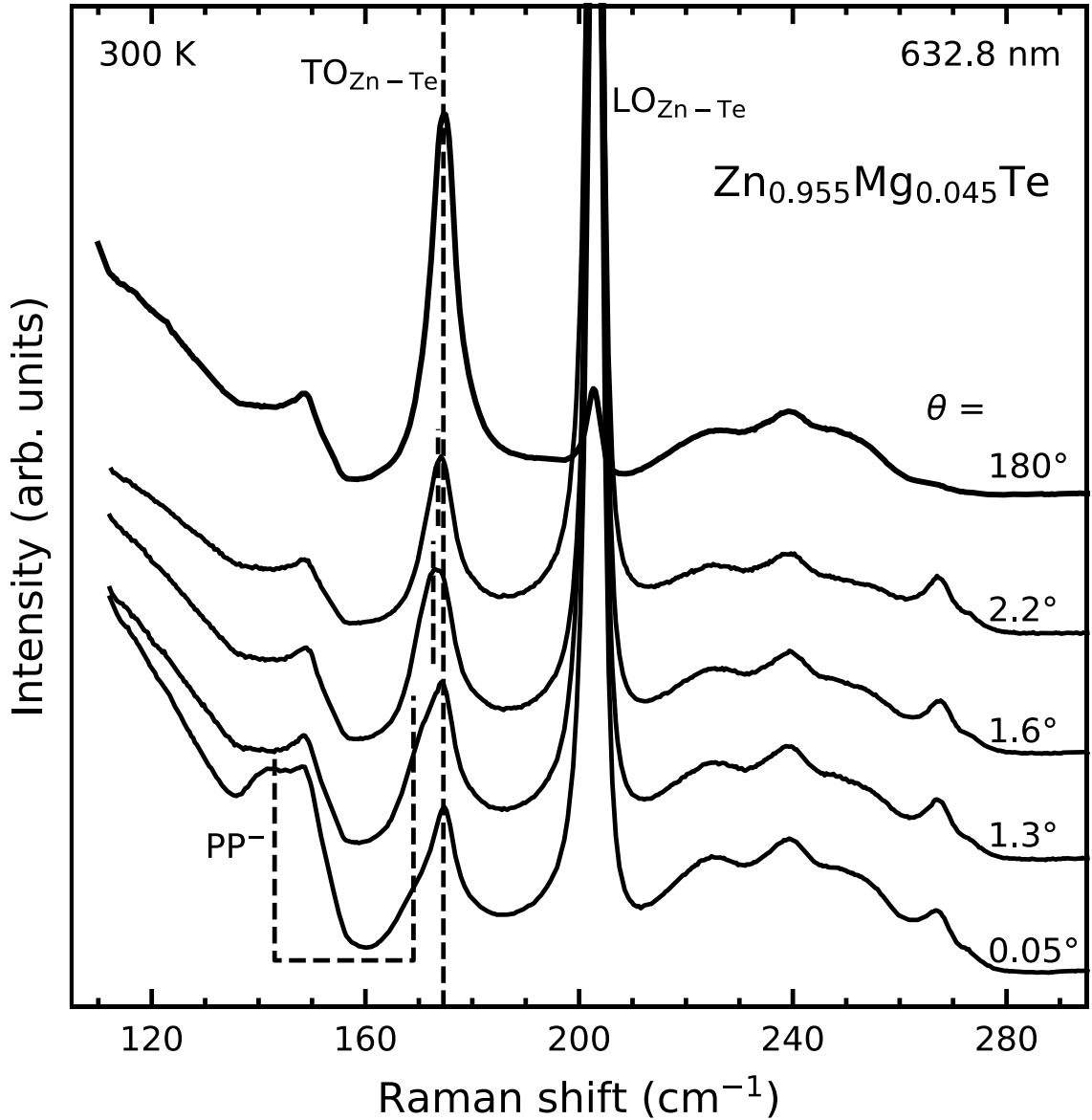

**FIG. S3. Zn$_{0.955}$Mg$_{0.045}$Te $\theta$-dependent Raman spectra.** Backward (thick curve) and near-forward (thin curves) Raman spectra of Zn$_{0.955}$Mg$_{0.0455}$Te taken at 300 K with the 632.8 nm laser excitation at various scattering angles ($\theta$), as indicated. The $\theta$-dependence of PP up to the native TO mode is emphasized (dashed curves).

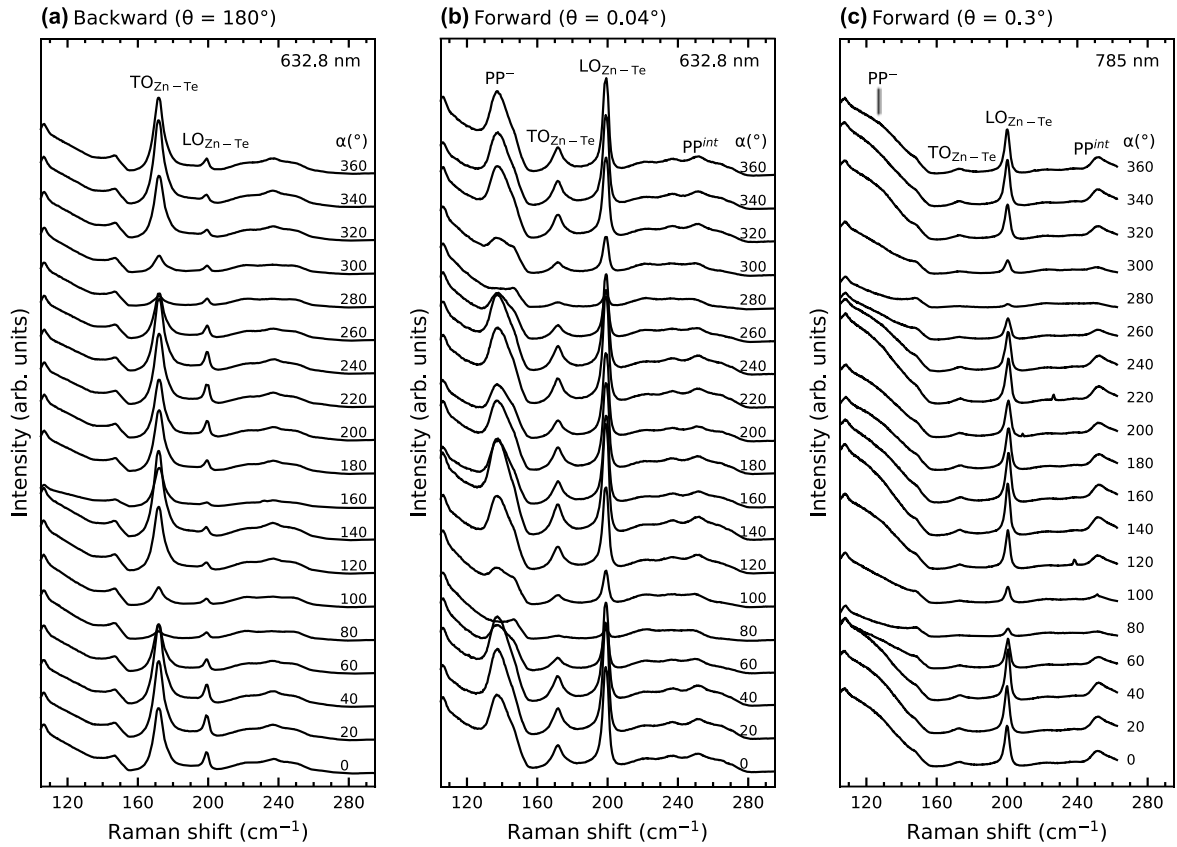

**FIG. S4. Zn$_{0.91}$Mg$_{0.09}$Te polarized Raman spectra.** (a) Zn$_{0.91}$Mg$_{0.09}$Te backward Raman spectra taken at minimal scattering angle ($\theta \sim 0$) with the 632.8 nm laser excitation in parallel polarizations rotated in pair at the sample surface over 360°. The angle $\alpha$ between polarizations and the [$\bar{1}10$] crystal axis, taken as an arbitrary reference, is specified. (b) and (c) Corresponding near-forward Raman spectra taken at the same sample spot with the 632.8 nm and 785.0 nm laser excitations, respectively. The exact $\theta$ angle is specified in each case.

SUPPLEMENTARY MATERIAL – ONLY REFERENCES